\documentclass[12pt]{article}

\usepackage{amsmath}
\usepackage{amssymb}
\usepackage{epsfig}
\usepackage{mathrsfs}
\usepackage{bm} 
\usepackage{cite}
\usepackage{enumitem}

\usepackage{setspace}
\usepackage[letterpaper, portrait, margin=2.5cm]{geometry}

\begin{document}

\title{
\bfseries
Frequency-domain nonlinear optics in two-dimensionally patterned quasi-phase-matching media}
%\date{}
\date{\vspace{-5ex}}
\maketitle

\author{C. R. Phillips$^{1,*,\dagger}$, B. W. Mayer$^{1,\dagger}$, L. Gallmann$^{1,2}$, and U. Keller$^1$}

\phantom{ }

\author{
\itshape
\noindent
$^1$: Department of Physics, Institute of Quantum Electronics, ETH Zurich, 8093 Zurich, Switzerland
\\
$^2$: Institute of Applied Physics, University of Bern, 3012 Bern, Switzerland
\\
$^\dagger$ These authors contributed equally
\\
$^*$: Corresponding author: cphillips@phys.ethz.ch}

\begin{abstract}
Advances in the amplification and manipulation of ultrashort laser pulses has led to revolutions in several areas. Examples include chirped pulse amplification for generating high peak-power lasers, power-scalable amplification techniques, pulse shaping via  modulation of spatially-dispersed laser pulses, and efficient frequency-mixing in quasi-phase-matched nonlinear crystals to access new spectral regions.  In this work, we introduce and demonstrate a new platform for nonlinear optics which has the potential to combine all of these separate functionalities (pulse amplification, frequency transfer, and pulse shaping) into a single monolithic device. Moreover, our approach simultaneously offers solutions to the performance-limiting issues in the conventionally-used techniques, and supports scaling in power and bandwidth of the laser source. The approach is based on two-dimensional patterning of quasi-phase-matching gratings combined with optical parametric interactions involving spatially dispersed laser pulses. Our proof of principle experiment demonstrates this new paradigm via mid-infrared optical parametric chirped pulse amplification of few-cycle pulses.
\end{abstract}

% ------------------------------------------------------------------- %
% ------------------------------------------------------------------- %
% Notes to self
% ------------------------------------------------------------------- %
% ------------------------------------------------------------------- %

% should we separate main-text reference list from references in methods, but keep common numbering for both lists?
% adding in abbreviations to reference list manaully [v4c for first version of this]
% using naturemag_edited [original file naturemag from https://www.ctan.org/tex-archive/macros/latex/contrib/nature?lang=en]
% do we have to put the figures at the end or in a separate file? One-page instructions say it's ok to have them in the main text
% ------------------------------------------------------------------- %
% latest numbers from checking in Word (copy paste from TeX, after removing numbered equations)
% 150 words [abstract]

% 495 words [intro] 
% 515 words [sec2, QPM device]
% 762 words [sec3, simulations]
% 542 words [sec4, experiment]
% 215 words [sec5, tandem]
% 441 words [conclusions]

% --> 2970 total

% ------------------------------------------------------------------- %
% ------------------------------------------------------------------- %

\section{Introduction}
\label{sec:sec_introduction}

Intense ultrashort laser pulses play a pivotal role in numerous areas of science and technology. In industry they have enabled  advances in micromachining, while in science they enable a broad class of intense light-matter interactions, with applications such as resolving attosecond dynamics in  atoms and molecules \cite{krausz_attosecond_2009, gallmann_attosecond_2012}, generation of soft x-ray radiation via high harmonic generation \cite{McPherson_multiphoton_1987, ferray_multiple-harmonic_1988}, and driving relativistic laser-plasma processes \cite{Mourou_relativistic_2006}. There is thus major interest in advancing these laser sources on several fronts, including wavelength (towards the mid-infrared), pulse duration (towards single-optical-cycle pulses), and repetition rate (to rapidly perform experiments and obtain good signal to noise ratios, while avoiding detrimental high-intensity effects). A compelling approach to generate the required sources is optical parametric chirped pulse amplification (OPCPA) \cite{witte_ultrafast_2012}, where an intense and narrow-bandwidth pump pulse amplifies a broad-band, temporally chirped signal pulse in a nonlinear crystal. In recent years there has been rapid growth in high-power ultrafast lasers \cite{russbueldt_compact_2010, eidam_femtosecond_2010, negel_kW_2013, saraceno_275_2012, saraceno_ultrafast_2014}. With OPCPA, the energy of these lasers can be transferred to few-optical-cycle pulses with user-chosen center wavelengths from the visible to far-infrared.

Nonetheless, optical parametric processes, including but not limited to OPCPA, present numerous challenges. Complicated non-collinear beam geometries and high laser intensities close to the damage threshold are needed to achieve phase-matched amplification over an ultra-broadband spectrum. Moreover, amplification occurs at all points in space and time of the pump pulse, and exhibits back-conversion of energy from the signal to the pump if the intensity is too high \cite{armstrong_interactions_1962}. Thus, maintaining the desired interaction across the spatial and temporal/spectral profiles of the interacting ultrashort waveforms remains a demanding problem. 

Recently, a complementary approach to OPCPA has been introduced, termed frequency domain optical parametric amplification (FOPA) \cite{chen_ultrabroadband_2010, schmidt_FOPA_2014}. The seed-pulse is dispersed spatially via a 4-f arrangement analogous to a pulse shaper \cite{weiner_pulse_shaping_2000}, with amplification occurring at the Fourier plane. By filling this plane with several birefringent phase-matching crystals placed side by side, as in the first experimental demonstration of the FOPA technique \cite{schmidt_FOPA_2014}, the phase-matching condition for different spectral regions can be adjusted separately, thereby relaxing one of the key constraints of conventional OPCPA systems. Moreover, because the seed is spatially chirped, its effective pulse duration can be matched to that of the few-ps pump pulse, allowing for efficient energy transfer. On the other hand, drawbacks to this powerful approach are that the optical path length through each crystal must be precisely matched, the complexity scales with the number of crystals used, and pre-pulses can be introduced by any diffraction from the edges of the crystals.

Here, we introduce and demonstrate a new paradigm for nonlinear-optical devices based on combining spatially dispersed laser pulses with a two-dimensionally patterned quasi-phase-matching (QPM) medium. This approach represents an extremely versatile yet experimentally simple platform, allowing for the limitations of existing nonlinear devices, including the above-mentioned drawbacks of the FOPA, to be overcome systematically by lithographic patterning of optimal 2D-QPM gratings. We experimentally demonstrate the approach with a mid-infrared FOPA.

\section{Two-dimensional quasi-phase-matching devices}
\label{sec:sec_2D_QPM_device}

\begin{figure*}[th!]
\centering
\includegraphics[width=1\textwidth]{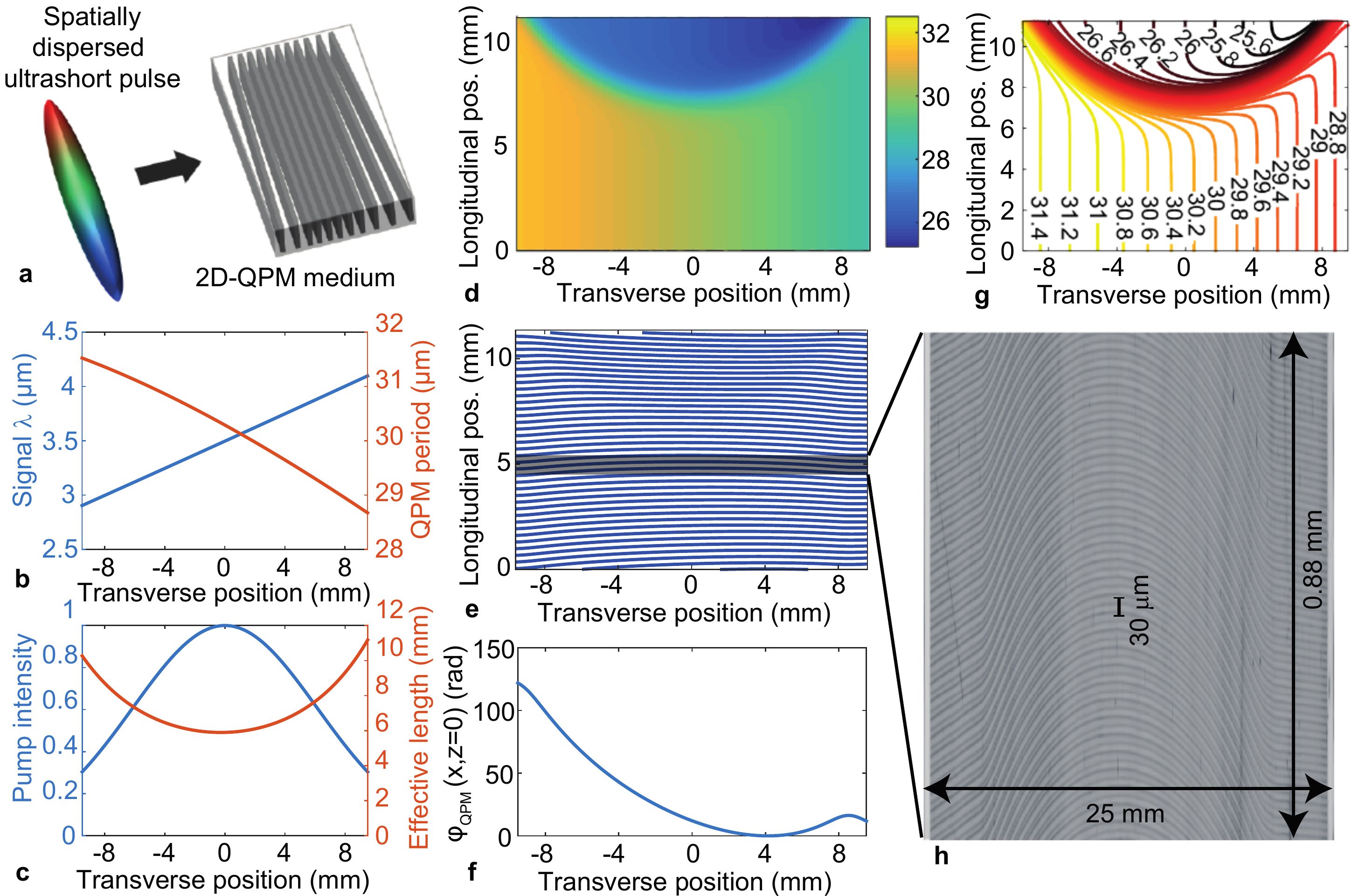}
\caption{
Illustration of 2D-QPM concept for frequency-domain optical parametric processes.
 (a) Schematic of the relevant experimental configurations, with a spatially chirped ultrashort pulse incident on a two dimensionally patterned QPM (2D-QPM) medium designed to individually address the different spectral components of the pulse. 
 (b) Signal wavelength versus transverse position in a 4-f pulse shaper (grating frequency 75~lines/mm, f=200~mm), and QPM period assuming a 1064-nm pump \cite{gayer_temperature_2008}. 
 (c) Example pump intensity profile across transverse position, and corresponding effective length to  achieve a flat small-signal gain profile.  
 (d) QPM period (in $\mu$m) obtained by combining (b) and (c), together with a smooth variation of the period along the longitudinal direction (see Methods). 
 (e) A selection of ferroelectric domain profiles (every 1/20$^{th}$ domain) for the QPM period mapping described by (d). 
 (f) Absolute QPM phase $\phi_{QPM}$ at the input position $z=0$ corresponding to part (e).
 (g) Contours of constant period corresponding to part (d).
 (h) Stretched image of the fabricated QPM grating in the 1-mm-thick MgO:LiNbO$_3$ crystal used for our experiments. The image was constructed via a series of microscope images of the $+z$ facet of the crystal along the transverse direction. The surface was etched to reveal the ferroelectric domain inversions.
}
\label{fig:fig1_device_concept}
\end{figure*}

In QPM \cite{armstrong_interactions_1962, fejer_quasi-phase-matched_1992}, the sign of the nonlinear coefficient is periodically or aperiodically inverted, augmenting the phase-matching condition with a term $K_g$ to yield $|\mathbf{k_p}-\mathbf{k_s}-\mathbf{k_i}-\mathbf{K_g}|\approx 0$, where $k_j$ are the wavevectors of the interacting waves. In periodically poled ferroelectric materials such as LiNbO$_3$, a lithography mask defines the QPM grating with high robustness \cite{fejer_quasi-phase-matched_1992, hum_QPM_2007, phillips_parametric_2013}. Thus, whereas birefringent phase-matching relies only on favorable material properties, QPM media can be freely engineered via lithography. For example, chirped QPM gratings can extend the phase-matching bandwidth well beyond that of periodic QPM gratings \cite{charbonneau-lefort_optical_2008, suchowski_octave-spanning_2013}. Here we take a much more general approach by using fully two-dimensional QPM (2D-QPM) patterns to tailor the nonlinear interactions experienced by spatially-separated spectral components, as illustrated conceptually in Figure \ref{fig:fig1_device_concept}(a). 

First, the QPM period can be varied in the transverse direction such that each spectral component is perfectly phase-matched. Figure \ref{fig:fig1_device_concept}(b) shows an example of this procedure for a mid-infrared pulse. In general, we can vary the QPM period continuously according to the exact trajectory required by material dispersion, with no inherent bandwidth constraint. Moreover, the linear properties of the QPM crystal remain homogeneous, so only a single, monolithic, plane-parallel crystal is required.

Next, figure \ref{fig:fig1_device_concept}(c) shows an example Gaussian intensity profile of the pump beam. Normally,  such changes in pump intensity would significantly change the gain for different spectral components. However, the 2D-QPM concept enables variation of the QPM grating properties \textit{along} the beam propagation direction as well. As an example to illustrate this general capability, we show in figure \ref{fig:fig1_device_concept}(c) how the effective length $L$ can be matched to the pump beam's intensity profile, thereby modifying the nonlinear interaction in order to flatten the small-signal gain spectrum.

To design a practical QPM grating profile meeting this criterion, we introduce a smooth, nonlinear variation in the QPM period. A corresponding map of QPM period is shown in figure \ref{fig:fig1_device_concept}(d). To fabricate this design, we introduce the absolute phase of the QPM structure, $\phi_{QPM}(x,z)$, given by 
\begin{align}
\label{eq:phi_QPM}
\phi_{QPM}(x,z)=\phi_{QPM}(x,0) + \int_0^z K_g(x,z')dz',
\end{align}
where the local QPM period is $2\pi/K_g$, and the input QPM phase $\phi_{QPM}(x,0)$ is a design degree of freedom (see Methods). Given $\phi_{QPM}$, the nonlinear coefficient satisfies $d(x,z)=\mathrm{sgn}(\cos(\phi_{QPM}(x,z)))$. The QPM phase is imparted to any waves generated during the nonlinear process: therefore, it introduces an opportunity to create an arbitrary phase mask for pulse shaping purposes. We discuss this potential in section \ref{sec:sec_tandem_concept}. 

The chosen $\phi_{QPM}(x,0)$ function is shown in figure \ref{fig:fig1_device_concept}(f), while figure \ref{fig:fig1_device_concept}(e) shows several of the resulting ferroelectric domains. We emphasize that the domains have no discontinuities, but have a significant curvature. This unique feature strongly contrasts with conventional mechanically-tunable QPM devices, which utilize multiple separate gratings for discrete tuning \cite{myers_multigrating_1996}, or straight but ``fanned'' ferroelectric domains in fan-out devices used for continuous tuning \cite{powers_continuous_1998}.  

To confirm that these domain profiles could be fabricated, we inspected the entire width of the devices used, as illustrated in figure \ref{fig:fig1_device_concept}(h), where we show a part of the successfully fabricated grating. The figure is stretched along the longitudinal direction to make the individual curved domains visible. There were no noticeable domain errors in the entire 1-mm-thick MgO:LiNbO$_3$ crystal (dimensions: 25-mm width by 12-mm length).

\section{Frequency-domain parametric interactions in 2D-QPM media}
\label{sec:sec_modeling}

\begin{figure*}[th!]
\centering
\includegraphics[width=1\textwidth]{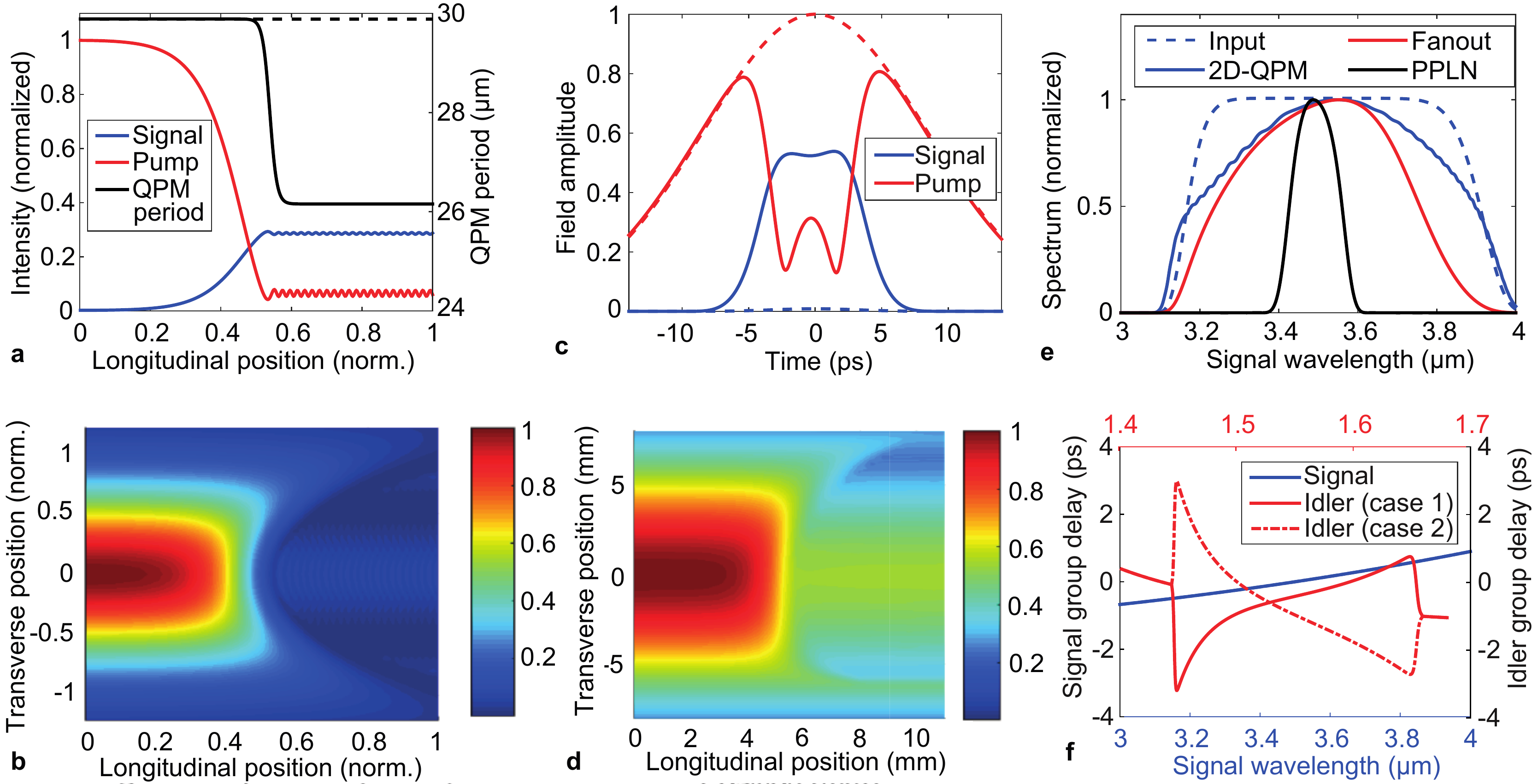}
\caption{
Modeling of frequency domain optical parametric amplification (FOPA) in a 2D-QPM medium. 
(a) Plane- and continuous-wave interaction in a longitudinally-varying QPM grating, showing the procedure used to switch off the parametric amplification after a certain distance through the crystal. 
(b) A series of simulations like (a), showing the evolution of the pump as a function of transverse position. At each transverse position, we perform a separate plane- and continuous wave simulation, with the pump intensity and effective length according to figure \ref{fig:fig1_device_concept}(c). 
(c) Full spatiotemporal simulation of the FOPA process. The figure shows the output electric field envelopes of the pump and signal for the transverse position $x=0$. 
(d) Evolution of the pump fluence through the crystal as a function of transverse position. 
(e) Output signal spectra for three cases: the 2D-QPM grating pattern introduced here; a simpler fanout pattern with no longitudinal variation; and the simplest case of a standard periodic grating. 
(f) Group delay spectra for the signal and idler, assuming the 2D-QPM grating pattern. For case 2, the pattern is flipped with respect to the longitudinal coordinate, changing the phase mask seen by the idler wave (derived in Supplementary section 3) but not changing the gain.
}
\label{fig:fig2_modeling}
\end{figure*}

We next elucidate the parametric process occurring in 2D-QPM media. We focus on 4-f pulse shaper arrangements as the means to introduce spatial chirp. Figure \ref{fig:fig2_modeling}(a) shows a simulation of a simplified situation involving a continuous-wave pump and signal, but including the longitudinal variation of the grating. Exponential amplification occurs, followed by depletion of the pump, followed by a rapid change of the QPM period to ``turn off'' the interaction. 

The capability to turn off or modify the parametric interaction in other ways within the nonlinear crystal is unique to structured QPM devices. Moreover, we show in figure \ref{fig:fig2_modeling}(b) how these modifications can be performed in a frequency-dependent fashion. The figure shows a series of the simulations from figure \ref{fig:fig2_modeling}(a) for different transverse positions: by turning off the interaction after the relevant effective length, back-conversion of energy to the pump is suppressed. Importantly, although frequency-dependent modifications can be accomplished in longitudinally-chirped QPM devices using one or multiple QPM gratings \cite{charbonneau-lefort_tandem_2005, charbonneau-lefort_optical_2008, phillips_design_2013}, such devices are ultimately limited by coupling between different parts of the spectrum \cite{phillips_design_2014}. In contrast, the spatial chirp in the FOPA decouples the spectral components more robustly, enabling greater flexibility.

To reveal the extent of this decoupling for the more subtle case involving real pulsed beams, the complete spatiotemporal profile of the signal must be considered. Our analysis in Supplementary section 2, where we derive this profile, reveals a correspondence between the spatial profile at the input diffraction grating of the 4-f setup and the temporal profile in the Fourier plane. Consequently, the duration of the spatially-chirped signal pulse is 
\begin{align}
\label{eq:tau_eff}
\tau_{eff}(\lambda)
%=\frac{\lambda^2}{\pi c|d\lambda/dx|w(\lambda)}
\approx \frac{\lambda}{c}\frac{\Delta x}{\Delta\lambda}\frac{w_{in}(\lambda)}{f},
\end{align}
where $\Delta\lambda$ is the range of wavelengths involved, and $\Delta x$ is the spatial extent of those wavelengths. As well as highlighting an important physical aspect of frequency-domain nonlinear optics, equation (\ref{eq:tau_eff}) allows comparison of $\tau_{eff}$ to the pump duration, which yields design guidelines for efficient operation.

To capture the nonlinear dynamics of the amplification process, we developed a numerical model for nonlinear mixing processes in 2D-QPM media (see Methods).  Figure \ref{fig:fig2_modeling}(c) shows the input and output of a full spatiotemporal simulation, accounting for propagation coordinate $z$, transverse dimension $x$, and time $t$ (2+1D). The input pump (dashed red) has duration 14~ps (FWHM). The effective input signal (dashed blue) has duration $\approx\!5.2$~ps (FWHM). The regions of the pump overlapped with the signal experience strong depletion, but since the signal is shorter than the pump in this example, the temporal wings of the pump remain undepleted. Figure \ref{fig:fig2_modeling}(d) shows how this issue manifests as a function of transverse beam position, by integrating over the time coordinate. In contrast to  \ref{fig:fig2_modeling}(b), complete depletion of the pump does not occur. 

The importance of a fully two-dimensional pattern is shown in figure \ref{fig:fig2_modeling}(e), which plots the simulated output spectrum for three cases. The solid blue curve corresponds to a 2D-QPM pattern [see figure \ref{fig:fig1_device_concept}(d)], showing amplification of the full input spectrum. The red curve uses a ``fanout'' grating [QPM period varied transverse to the beam according to figure \ref{fig:fig1_device_concept}(b)] but has no longitudinal variation. There is a substantial reduction of the bandwidth, and over-driving the device to recover the lost bandwidth introduces strong spatiotemporal distortions. Finally, the black curve models a simple periodic grating: in this case, there is a drastic reduction in bandwidth.

Beyond amplifying the interacting waves, the QPM device can act as an arbitrary phase mask, thereby offering a unique platform for simultaneous gain and pulse shaping. This shaping is provided by the QPM phase $\phi_{QPM}$ [figure \ref{fig:fig1_device_concept}(f)], which is imparted to the generated wave. The idler spectral phase $\phi_i(\nu)$  can be approximated as
\begin{align}
\label{eq:phi_i_spectrum}
\phi_i(\nu)
\propto 
-\phi_s(\nu_p-\nu) + \phi_{QPM}(x_i(\nu),0) - k_i(\nu)L,
 \end{align} 
where $k_i(\nu)$ is the idler wavevector and $\phi_j(\nu)$ is the spectral phase of wave $j$. A more general expression for $\phi_i(\nu)$, including additional terms to account for the longitudinal variation of the QPM grating, is given in Supplementary section 3.

There is significant freedom in choosing $\phi_{QPM}(x,0)$, subject only to constraints on the ferroelectric domain angles that can be fabricated. Consequently, very large phases can be imparted onto the idler. Unlike conventional pulse shapers, this phase is fully continuous, and no discrete wrapping between $0$ and $2\pi$ phase is needed. These properties are illustrated in figure \ref{fig:fig2_modeling}(f), which shows the group delay spectra of the signal and idler for two cases. The signal group delay is determined by the dispersion of the material, while the idler group delay is determined by equation (\ref{eq:phi_i_spectrum}). The two cases shown correspond to two orientations of a particular QPM grating, thereby emphasizing how the group delay can be modified substantially, over several picoseconds or more, without altering the amplification characteristics.

\section{Experiment}
\label{sec:sec_experiment}

\begin{figure*}[h!t]
\centering
\includegraphics[width=1\textwidth]{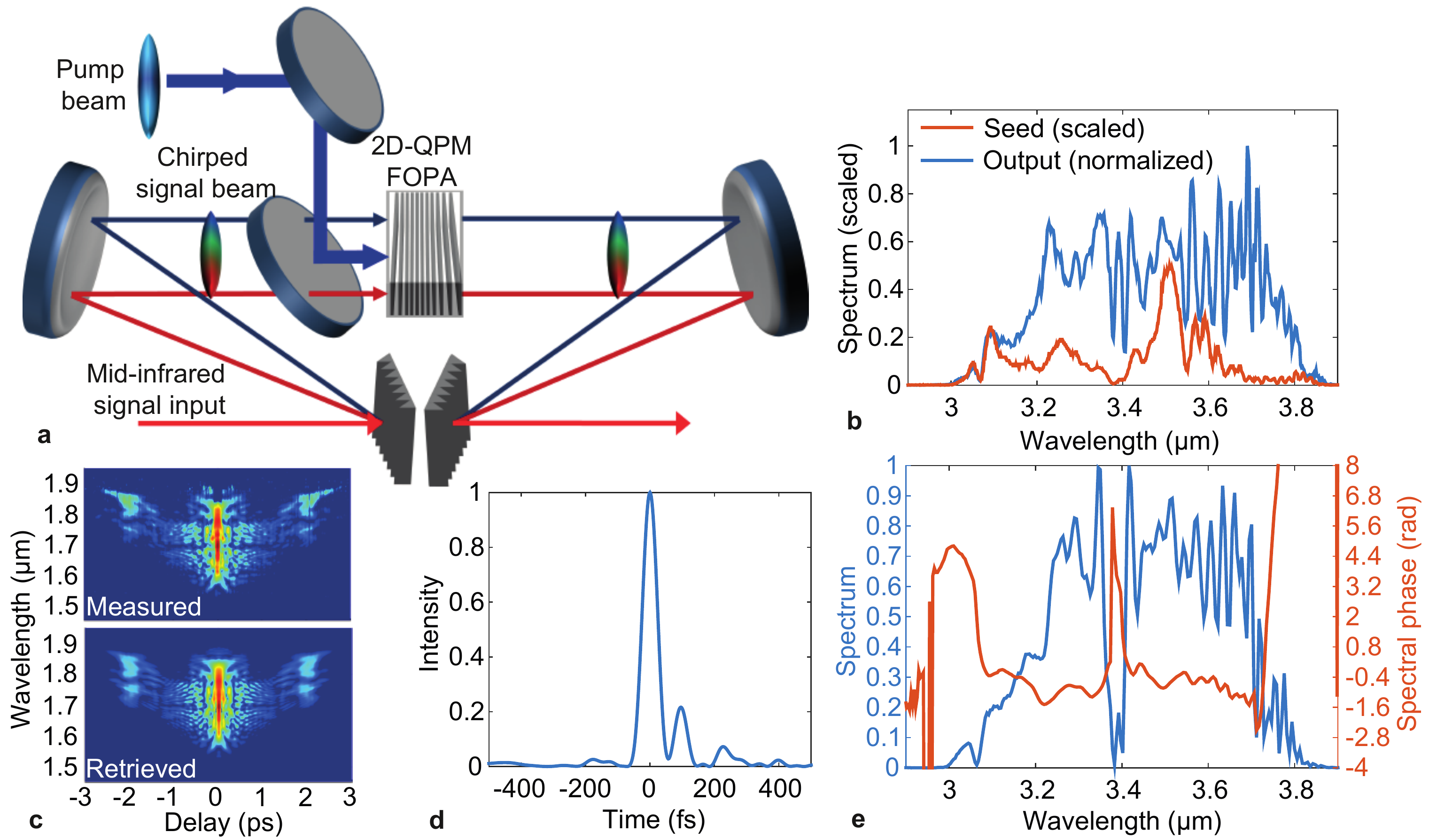}
\caption{
Experimental setup and results. (a) Schematic of the experimental FOPA setup. (b) Input and output spectra; the output spectrum is normalized, while the seed spectrum is scaled so that it is visible on the same scale as the pump. (c) Measured and retrieved SHG-FROG spectrograms. (d) Reconstructed pulse profile. (e) Reconstructed spectrum and phase.
}
\label{fig:fig3_experiment}
\end{figure*}

To demonstrate the technique, we implemented the 2D-QPM FOPA as the final stage of a mid-infrared OPCPA system \cite{mayer_OL_2013, mayer_achromatic_2014} (see Methods). A schematic of the FOPA is shown in figure \ref{fig:fig3_experiment}(a). The diffraction gratings have $75$~lines/mm, designed for a blaze wavelength of 4000~nm. The focal length of the 4-f arrangement is $f=200$~mm. The elliptical pump beam is collinearly overlapped with the spatially-chirped mid-infrared signal by a dichroic mirror which transmits the signal and reflects the pump. An identical mirror removes the remaining pump after the 2D-QPM crystal (not shown). 

The pump laser has a duration of $\approx\!14$~ps, and an average power of 16.5~W at a repetition rate of 50~kHz. The beam $1/e^2$ full-width in the QPM crystal is $\approx\!27.4$~mm in the horizontal and $\approx\!140$~$\mu$m in the vertical. The mid-infrared seed has an average power of 6.4~mW before the first diffraction grating. Its spectral components are spatially chirped according to figure \ref{fig:fig1_device_concept}(b). To improve the temporal overlap with the pump pulses according to equation (\ref{eq:tau_eff}), we use a cylindrical telescope prior to the 4-f arrangement to obtain a relatively large $1/e^2$ beam size of $\approx 12$~mm ($1/e^2$ full width) in the horizontal direction.

Measured spectra are shown in figure \ref{fig:fig3_experiment}(b).  The compressed output power after the second diffraction grating was 1.03~W, corresponding to 20.6~$\mu$J pulse energy. Accounting for losses of the diffraction grating ($\approx\!31$~$\%$), the dichroic mirror ($\approx\! 5\%$), and beam-routing optics, we estimate an average power of $\approx\!1.65$~W directly after the antireflection-coated 2D-QPM crystal, corresponding to 33~$\mu$J pulse energy. We thus infer a quantum efficiency (ratio between output signal photons and input pump photons) of 32\%, which is a substantial improvement over our previous OPCPA configuration based on non-collinear power amplification in a conventional PPLN crystal \cite{mayer_achromatic_2014}. 

To compress the output pulses, we adjusted the 1550-nm seed pulses in the OPCPA front-end. The compressed pulses were measured using second-harmonic generation frequency resolved optical gating. The measured and retrieved spectrograms [figure \ref{fig:fig3_experiment}(c)] exhibit good agreement. The reconstructed pulse profile is plotted in figure \ref{fig:fig3_experiment}(c), indicating compression to 53~fs (FWHM); this is mainly limited by the available seed bandwidth (43~fs transform limit, corresponding to four optical cycles). The reconstructed spectrum and phase are shown in figure \ref{fig:fig3_experiment}(e), in good agreement with the independently-measured spectrum [figure \ref{fig:fig3_experiment}(b)]. The fluctuations on the spectra can be explained by considering the spectral broadening of our 1550-nm seed in the OPCPA front-end (see Supplementary section 5). 

Our proof of principle experimental results establish the viability of the 2D-QPM FOPA technique. The demonstrated conversion efficiency already exceeds the state of the art for mid-infrared systems \cite{moses_highly_stable_2009,deng_carrier_2012,hemmer_18uJ_COL_2013,hong_multi-mJ_2014,mayer_achromatic_2014}. Nonetheless, the comparatively short duration of the signal in the Fourier plane limited the achievable  efficiency somewhat. Based on equation (\ref{eq:tau_eff}), using $w_{in}(\lambda)=12$~mm ($1/e^2$ full-width), we estimate the signal pulse duration as 6.35~ps (full-width at half-maximum), which should be compared to the 14~ps pump. As shown in figure \ref{fig:fig2_modeling}(d), this mismatch in pulse durations reduces the efficiency compared to the simplified continuous wave case of figure \ref{fig:fig2_modeling}(b). Note, however, that since no poling errors were present in the fabricated device [figure \ref{fig:fig1_device_concept}(h)], QPM gratings with significantly larger widths up to $\sim\!60$~mm are feasible (limited by wafer width). Scaling $\Delta x$ in equation (\ref{eq:tau_eff}) accordingly would yield a corresponding increase in signal pulse duration.

\section{Dual-grating concept for signal shaping}
\label{sec:sec_tandem_concept}

\begin{figure}[tb]
\centering
\includegraphics[width=0.5\linewidth]{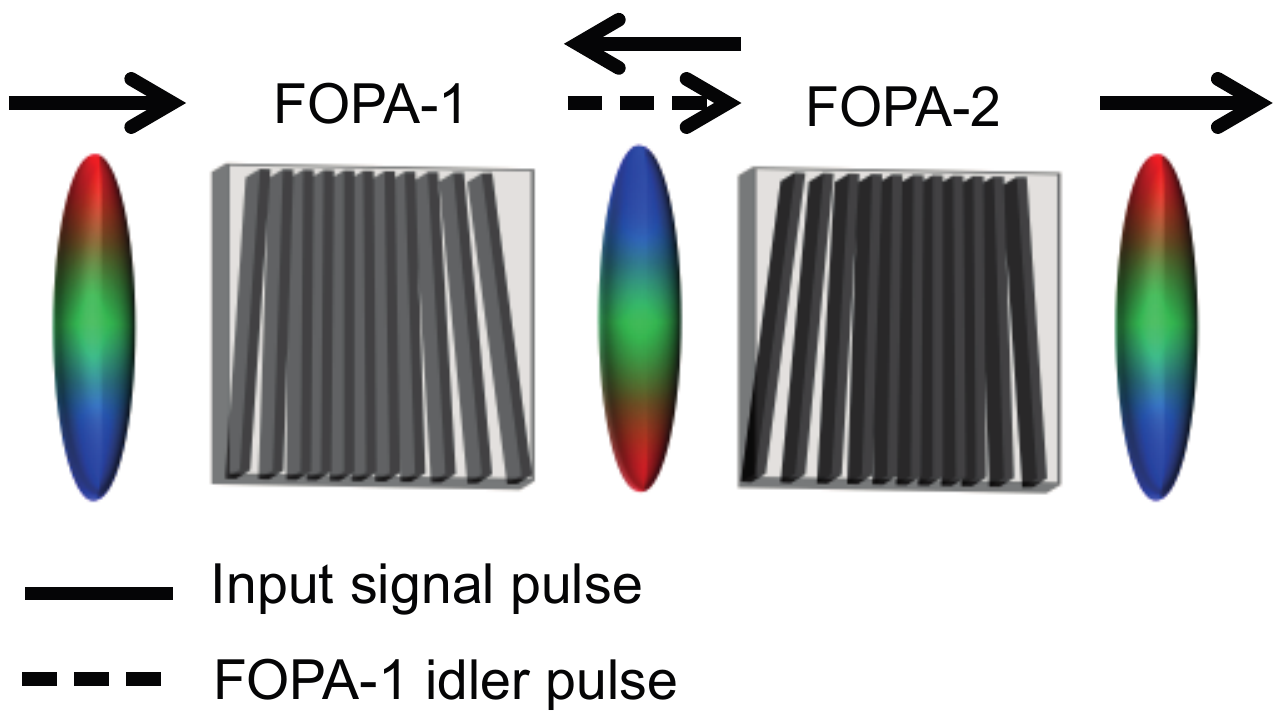}
\caption{
Concept for simultaneous amplification and pulse shaping of the input signal pulse. Two FOPA crystals are used. The idler generated in the first crystal seeds the second crystal. In this way, the output of the second crystal inherits the QPM phases of FOPA-1 and FOPA-2. These QPM phases can be engineered to accomplish the desired pulse shaping functionality, for example compensation of higher order spectral phase terms of the signal.
}
\label{fig:fig4_tandem_concept}
\end{figure}

In this section we propose an extension of the demonstrated approach that enables arbitrary shaping of the signal pulse. This is an important point because, unlike the signal, the spatial chirp of the generated idler is not linear in wavelength, and hence cannot be fully removed by a diffraction grating. The key concept, depicted in  figure \ref{fig:fig4_tandem_concept}, is to have a second crystal seeded by the idler wave from the first crystal. 

We assume an initial signal input to crystal FOPA-1 with center frequency $\nu$. In FOPA-1, an idler wave (frequency $\nu_p-\nu$) is generated. The spectral phase of this idler is related to the QPM phase of FOPA-1 according to equation (\ref{eq:phi_i_spectrum}). Next, the FOPA-1 output signal is discarded, while the FOPA-1 idler (frequency $\nu_p-\nu$) seeds FOPA-2. The idler wave of FOPA-2 is at the original frequency $\nu$, and its phase is related to the QPM phase of FOPA-2. Therefore, the final signal output at frequency $\nu$ of the crystal pair has experienced a phase modulation according to the difference of the QPM phases of the two crystals (derived in Supplementary section 3). By adjusting the QPM phases by design of the crystals, an arbitrary effective phase mask for the signal wave can be achieved, thereby combining amplitude and phase shaping into a single crystal-pair.

\section{Conclusions}
\label{sec:sec_conclusions}

In conclusion, by combining a spatially chirped input wave with a two-dimensionally patterned QPM crystal, we have introduced and demonstrated a new platform for nonlinear optics that has unprecedented flexibility and overcomes the limitations and trade-offs inherent in conventional devices. By using QPM crystals with curved domains fabricated with high fidelity, we have shown that the technique is scalable in bandwidth, since the QPM period trajectory in the crystal can be matched to the input wave, even for extremely broad bandwidths where the QPM period changes significantly and nonlinearly versus position. The approach is applicable to a wide variety of nonlinear-optical devices, including harmonic generation and optical parametric amplification. 

In contrast to conventional ultrafast processes, the 2D-QPM FOPA consists of a continuum of narrow-band OPA interactions across the transverse dimension of the crystal, with the freedom to adjust the character of these interactions across the spectrum via the QPM pattern. For our proof of principle experiment, we combined a transverse variation in QPM period with a longitudinal variation to compensate for the pump beam shape, thereby addressing two ubiquitous problems in nonlinear optics (phase-matching, and use of non-flat-top beams). Moreover, a broad class of such longitudinal variations are possible: examples include introducing a longitudinal chirp in the QPM period, or introducing a second QPM segment for additional functionality such as harmonic generation. Further exploration of these capabilities, as well as optimization of the experimental parameters, provides a rich framework for future work. Pump beam shaping, as in \cite{schmidt_FOPA_2014}, could help enable such exploration by reducing constraints due to the pump spatial profile.

Beyond amplitude shaping and bandwidth scaling, we showed in section \ref{sec:sec_tandem_concept} how the FOPA can act as an arbitrary phase mask as well, thereby combining the key functionalities of broadband amplification and pulse shaping in a single crystal-pair. This capability should help greatly simplify nonlinear-optical systems, since all of the challenging phase-matching, amplification, and dispersion management aspects can be lumped into a single, highly engineerable component. The slab-like geometry of the FOPA makes it power-scalable, since it supports one-dimensional heat flow along the thin axis of the crystal. The advent of large-aperture QPM crystals provides great promise for energy scaling as well \cite{ishizuki_high-energy_2005, ishizuki_half-joule_2012, hum_quasi-phase-matched_2007}. Moreover, QPM media are available in diverse spectral ranges, for example covering from the ultraviolet via LBGO \cite{hirohashi_lbgo_CLEO_2015}, to the far-infrared via orientation patterned GaAs and GaP \cite{lynch_growth_2008, schunemann_GaP_CLEO_2014}. Therefore 2D-QPM devices will be applicable for pulse generation, shaping, and amplification across the optical spectrum, from the deep-ultraviolet to far-infrared. We therefore expect that frequency-domain processes enabled by such 2D-QPM media will have broad impact and appeal for many areas of photonics.

% --------------------------------------------- %
% --------------------------------------------- %

\section*{Methods}

% --------------------------------------------- %
% QPM grating design
% --------------------------------------------- %

The 2D-QPM grating introduced in section \ref{sec:sec_2D_QPM_device} was designed by combining several techniques. To obtain the transverse variation of the QPM period (direction $x$), we calculate the position of the spatially chirped spectral components using the grating equation, and apply the Sellmeier relation to find material phase-mismatch $\Delta k_0=k_p-k_s-k_i$ \cite{gayer_temperature_2008} according to figure \ref{fig:fig1_device_concept}(b). For the longitudinal variation (direction $z$), we first determine the required effective length according to figure \ref{fig:fig1_device_concept}(c), and then apply a $z$-dependent offset in $K_g$; we construct the offset using hyperbolic tangent functions, in analogy to apodization profiles discussed in the context of chirped QPM media \cite{charbonneau-lefort_optical_2008, phillips_apodization_2013, phillips_femtosecond_2015}. The QPM profile for a particular $x$-slice is shown in figure \ref{fig:fig2_modeling}(a), black curve.

Rather than simply removing the QPM grating entirely in the switched-off regions, we maintain a 50-\% QPM duty cycle through the whole crystal, and instead switch off the amplification process by rapidly changing the QPM period away from phase-matching (i.e. a ``nonlinear chirp'' profile). Such a 50-\% duty cycle helps suppress photorefractive effects \cite{taya_photorefractive_1996, phillips_continuous_2011, schwesyg_pyroelectrically_2011}, in addition to the suppression already provided by MgO-doping of the crystal.

The appropriateness of the chosen QPM designs was tested with simulations of the OPA process including pump depletion. The value of $K_g$ at the end of the grating is not directly coupled to the OPA gain provided the phase-mismatch $\Delta k=k_p-k_s-k_i-K_g$ is made large enough. However, nonlinear phase shifts can be introduced by phase-mismatched nonlinear processes \cite{stegeman_cascading_review_1996}: 2D-QPM creates the opportunity to select the phase-mismatch to manipulate these nonlinear phase shifts. Accordingly, we selected the $K_g$ profile to yield phase shifts comparable in magnitude to those from the intrinsic $\chi^{(3)}$ nonlinearity \cite{phillips_opex_SC}. This approach, motivated by our recent work on adiabatic excitation of quadratic solitons \cite{phillips_femtosecond_2015}, is another unique capability of 2D-QPM media which we will explore in more detail in future work.

With respect to fabrication, our choice of $\phi_{QPM}(x,0)$ in equation (\ref{eq:phi_QPM}) minimizes the angles of the ferroelectric domains. These domain angles satisfy $\tan(\theta(x,y))=(\partial\phi_{QPM}/\partial x) / (\partial\phi_{QPM}/\partial z)$, and $\phi_{QPM}(x,0)$ allows the profile of these domain angles to be manipulated without altering the longitudinal derivative $\partial\phi_{QPM}/\partial z$ which determines the phase-matching properties. Lastly, given $\phi_{QPM}$, the lengths of the pattern on the lithography mask are offset from the desired 50-\% duty cycle QPM grating to account for domain spreading during poling of the ferroelectric domains.

% --------------------------------------------- %
% numerical simulations
% --------------------------------------------- %

For our numerical simulations, shown in figure \ref{fig:fig2_modeling}, we developed a unidirectional coupled-envelope model designed to model nonlinear mixing between the envelopes due to second- and third-order nonlinearities. The model allows in principle for arbitrary mixing between the nominal pump, signal, and idler envelopes as well as additional envelopes corresponding to sum- and difference-frequencies between the nominal envelopes. The model also supports arbitrary QPM media, provided the assumption of unidirectionality and paraxial diffraction hold. The spatial chirp of the broadband signal, and the full 2D-QPM phase $\phi_{QPM}(x,z)$, are included. The simulation in figure \ref{fig:fig2_modeling}(c) uses a ``2+1D'' model (transverse coordinate $x$, propagation coordinate $z$, and time $t$). Further discussion is given in Supplementary section 4.

% --------------------------------------------- %
% OPCPA front-end
% --------------------------------------------- %

Our experiments described in section \ref{sec:sec_experiment} use a mid-infrared OPCPA front-end containing two OPCPA pre-amplifiers \cite{mayer_OL_2013, mayer_achromatic_2014}. A schematic of the system is shown in Supplementary figure 1. The system uses two synchronized lasers for pumping and seeding the pre-amplifiers. The pump laser has a wavelength of 1064~nm and produces 14-ps pulses (FWHM) at a 50-kHz repetition rate, with 8~W average power. Approximately 5~W is directed to the pre-amplifiers, while the remaining power is directed to a home-built Innoslab-type amplifier, the output of which is used to pump the frequency domain OPA illustrated in figure \ref{fig:fig3_experiment}(a). The seed laser is a femtosecond fiber laser with subsequent erbium-doped fiber amplifiers. The laser has a wavelength of 1550~nm and produces 70-fs pulses at an 82~MHz repetition rate, with 250-mW average power. The seed pulses are first spectrally broadened in a dispersion shifted fiber (DCF3, Thorlabs) before being chirped in time with a silicon prism pair and 4-f pulse shaper arrangement. This chirp is transferred to the mid-infrared by the second pre-amplifier, and as such we can optimize the compression of the final amplified mid-infrared pulses by adjusting the dispersion of the infrared seed pulses.

The OPCPA pre-amplifiers are based on longitudinally chirped quasi-phase-matching gratings, implemented in MgO:LiNbO$_3$. We use the shorthand aperiodically poled lithium niobate (APPLN) to refer to them in \cite{mayer_OL_2013}. Both crystals have the same QPM design, and the pump, signal, and idler beams are all collinearly aligned in the crystals. After the first APPLN crystal, we discard the idler wave (mid-infrared output), and route the pump and signal outputs to the second crystal. After the second APPLN crystal, the pump and signal waves are discarded, and we extract the 3400-nm mid-infrared wave to seed the final amplifier (the FOPA). Further information on the system is given in Supplementary section 1.

The chirped QPM gratings used for the pre-amplifiers can in principle be scaled in bandwidth, but careful consideration must be given to several design constraints, described in detail in \cite{phillips_design_2014}. These constraints, which relate to favoring the desired OPA process over various unwanted processes, become restrictive when operating in the highly pump depleted regime corresponding to adiabatic frequency conversion. Therefore, the combination of longitudinally chirped QPM devices for convenient and alignment-insensitive pre-amplification to moderate energy levels, followed by the 2D-QPM FOPA for final power amplification, represents a compelling system arrangement which preserves bandwidth scalability, avoids the challenging parasitic processes of highly-saturated chirped QPM devices, and keeps complexity at a minimum since only one FOPA arrangement is required.

% --------------------------------------------- %

%\bibliographystyle{naturemag_edited}
%\bibliography{C:/Users/Chris/Dropbox/_References/phillips-refs-all}
%\bibliography{E:/Dropbox/_References/phillips-refs-all}

% --------------------------------------------- %
% --------------------------------------------- %

\section*{Funding Information}

This research was supported by the Swiss National Science Foundation (SNSF) through grants \#200020\_144365/1 and \#200021\_159975/1, and by Marie Curie International Incoming fellowship grant PIIF-GA-2012-330562 within the 7$^{th}$ European Community Framework Programme.

%\bigskip \noindent See \href{link}{Supplement 1} for supporting content.

\section*{Author Contributions}

C. R. P. and B. W. M. conceived, designed, and carried out the experiments. C. R. P. developed the theory and simulations, and wrote the manuscript. L. G. and U. K obtained the funding. All authors contributed to discussions of the results and the manuscript.

\end{document}